# Missile Attitude Control via a Hybrid LQG-LTR-LQI Control Scheme with Optimum Weight Selection

Saptarshi Das[1,2]
1. School of Electronics and Computer Science, University of Southampton, Southampton SO17 1BJ, UK.
Email: s.das@soton.ac.uk, saptarshi@pe.jusl.ac.in

Kaushik Halder[2]
2. Department of Power Engineering, Jadavpur University, Salt-Lake Campus, LB-8, Sector 3, Kolkata-700098, India.
Email: khalder.pe@research.jusl.ac.in

*Abstract*—This paper proposes a new strategy for missile attitude control using a hybridization of Linear Quadratic Gaussian (LQG), Loop Transfer Recovery (LTR), and Linear Quadratic Integral (LQI) control techniques. The LQG control design is carried out in two steps i.e. firstly applying Kalman filter for state estimation in noisy environment and then using the estimated states for an optimal state feedback control via Linear Quadratic Regulator (LQR). As further steps of performance improvement of the missile attitude control system, the LTR and LQI schemes are applied to increase the stability margins and guarantee set-point tracking performance respectively, while also preserving the optimality of the LQG. The weighting matrix ($Q$) and weighting factor ($R$) of LQG and the LTR fictitious noise coefficient ($q$) are tuned using Nelder-Mead Simplex optimization technique to make the closed-loop system act faster. Simulations are given to illustrate the validity of the proposed technique.

*Keywords*—Attitude control; canard missile; LQG; LQI; LQR; LTR; optimal control; optimum weight selection

## I. INTRODUCTION

Guidance and control law design for aerospace engineering and strategic applications has become a celebrated research topic in past few decades with its wide applicability both in industry and academia [1]-[2]. In this paper, a new control strategy for a canard configured missile is proposed using a hybrid approach of LQG, LTR and LQI schemes. The hybrid scheme retains the advantage of all these three individual control concepts and outperforms each of them for a step-command attitude control problem of canard missile. The canard configuration is widely used in homing missiles [3]-[4] which motivates us to study the control problem of this particular configuration [5]. The canard is basically a small wing of an aircraft. In this configuration of the missile, it is placed in front of the center of gravity of the missile to provide addition wing lift for flexible maneuvering. But this makes the overall control system loose internal stability due the non-minimum phase behavior i.e. having an open loop zero in the right-half *s*-plane. The system under study, being both controllable and observable [6], the task is now to design an efficient control scheme to track a step-input angle of attack command with stochastic observer based state-feedback control by varying the control inputs that directly affects the canard's deflection. Control of canard missile has been previously studied using sliding mode [6] and LQG [7] controller.

There are several attempts by contemporary researchers to design efficient control scheme for similar missile and guidance systems, using various techniques from control theory e.g. robust control [8], model predictive control [9], LQR [10], $H_\infty$ [11], sliding mode [12]-[14], fuzzy logic [15], particle filter [16], nonlinear control [17] etc. Among several approaches the LQG technique which is composed of optimal state estimation in the presence of process noise and measurement noise, followed by state-feedback regulator design using the observed states are quite popular [18]-[19]. The stochastic observer based state feedback control law can easily be implemented as an output feedback controller which is commonly known as the observer based controller [18]-[21]. The applicability of classical optimal and robust control theory like LQR, $H_2/H_\infty$ in aerospace application is restricted due to their deterministic nature of system analysis and design. Also, several constraints are imposed due to the high level of noisy observation of the practical system's input/output and unavailability of direct measurements for all the state variables. This motivates the use of Kalman filter or stochastic observer in such applications.

The disadvantage of Kalman filter based LQR design or LQG control is that it makes the open-loop system have very low stability margins (gain and phase margin) which can be further improved using a technique known as LTR by increasing a fictitious noise coefficient ($q$) in the process-noise covariance matrix in the Kalman filter design. This makes the open-loop system asymptotically approach the response obtained using LQR which is popular in control engineering practice to alleviate the ill-effects of LQG design [22]. The only constraint using LQG/LTR technique is that choosing a high value of $q$ makes the control signal large which might result in actuator saturation. Even with the LQG/LTR design, the set-point tracking performance is not guaranteed. The tracking can be enforced by externally tuning the LQR weights using an optimization algorithm but the tracking cannot be guaranteed under parametric variation in the system matrices unless the controller contains a built-in proportional-integral-derivative (PID) type scheme or its variants like a simple PI or at least an integral control [23]-[24]. To overcome the tracking problem in state-feedback control, the LQI scheme is proposed which considers the presence of an integrator in the forward path and acts as an output feedback controller beside the state-feedback controller implemented on either directly measured or observed states [25]-[27]. As a summary, the missile attitude control scheme must have a state-observer and state-feedback control (LQG), improved stability margin (LTR) and guaranteed tracking (LQI). The proposed LQG/LTR optimum weight selection and its fusion with LQI can be considered as a unification of important but discretely available concepts in the literature *viz.* LQG-PI tracking control weight tuning [28], LTR integral control [29] and LQG weight selection [30].

The rest of the paper is organized as follows: section II outlines the missile attitude control system and proposes the hybrid LQG-LTR-LQI control scheme. Simulation results with classical LQG/LTR design have been reported in section III. The proposed LQG-LTR-LQI control results have been shown in section IV with optimal choice of $Q$, $R$ and $q$. The paper ends with conclusions in section V, followed by the references.

## II. MISSILE ATTITUDE CONTROL SYSTEM DESCRIPTION AND PROPOSING A NEW HYBRID LQG-LTR-LQI SCHEME

### A. Missile Model in Canard Configuration

The linearized model of canard-configured missile [6] is described by the following state-space model in (1).

$$\begin{bmatrix} \dot{\alpha}(t) \\ \dot{q}(t) \\ \dot{\delta}(t) \end{bmatrix} = \begin{bmatrix} Z_\alpha & 1 & Z_\delta \\ M_\alpha + \Delta M_\alpha & M_q & M_\delta \\ 0 & h/\tau_s & -1/\tau_s \end{bmatrix} \begin{bmatrix} \alpha \\ q \\ \delta \end{bmatrix} + \begin{bmatrix} 0 \\ 0 \\ 1/\tau_s \end{bmatrix} u + \begin{bmatrix} 0 \\ 0 \\ 1 \end{bmatrix} \xi(t) \quad (1)$$

$$y(t) = \begin{bmatrix} 1 & 0 & 0 \end{bmatrix} \begin{bmatrix} \alpha & q & \delta \end{bmatrix}^T + \theta(t)$$

Here, the state variables are considered as the angle of attack ($\alpha$), pitch rate ($q$), canard's deflection ($\delta$), while $\{u, y\}$ are the control input and system output respectively. Other system parameters like seeker time constant ($\tau_s$), radome bore-sight error slope ($h$), different aerodynamic derivatives $\{Z_\alpha, Z_\beta, M_\alpha, M_q, M_\delta\}$ also affects the missile dynamics along with the unknown (but bounded) disturbances $\{\xi, \theta\}$. In the canard-configured missile, typical parameters for the aerodynamic derivatives and model perturbation are given by:

$$Z_\alpha = -2.7 \text{ s}^{-1}, Z_\delta = 0.27 \text{ s}^{-1}, M_\alpha = -5.5 \text{ s}^{-2}, M_q = -0.4 \text{ s}^{-1}, \quad (2)$$
$$M_\delta = -19 \text{ s}^{-2}, \tau_s = 0.05 \text{ s}, -1 \leq \Delta M_\alpha \leq 1 \text{ s}^{-2}, 0 \leq h \leq 0.1 \text{ s}$$

### B. LQG as a Combination of Kalman Filtering and LQR

Comparing the missile model (1) with the standard state-space model (3), one can identify system matrices (state, input, output, feed-through matrices given by $\{A, B, C, D\}$) and the matrix associated with the process-noise term ($\Gamma$). The process noise and measurement noise terms $\{\xi(t), \theta(t)\}$ are assumed to be zero mean Gaussian with covariance matrices ($\Xi, \Theta$) and also statistically independent.

$$\begin{aligned} x(t) &= Ax(t) + Bu(t) + \Gamma \xi(t) \\ y(t) &= Cx(t) + Du(t) + \eta(t) \end{aligned} \quad (3)$$

The task is to control the missile such that it tracks a step-input angle of attack command, minimising the tracking error and rejecting the unknown disturbances.

$$\begin{aligned} E[\xi(t)] &= E[\theta(t)] = 0, \quad E[\xi(t)\theta^T(t)] = 0 \\ E[\xi(t)\xi^T(t)] &= \Xi \geq 0, \quad E[\theta(t)\theta^T(t)] = \Theta > 0 \end{aligned} \quad (4)$$

In LQG design there are two steps *viz.* state estimation, followed by LQ optimal state feedback. In the estimation phase, the task is to find out the optimal state estimate $\hat{x}(t)$ that minimizes the covariance $E\left[(x-\hat{x})(x-\hat{x})^T\right]$. The optimal estimate is achieved using Kalman filter, where the Kalman gain ($K_f$) can be computed from (5) using the symmetric positive semi-definite matrix solution ($P_f = P_f^T$) of Algebraic Riccati Equation (ARE) reported in (6). Here in (5)-(6), subscript "$f$" denotes the Kalman *filter* to distinguish with similar ARE encountered in LQR *controller* (subscript "$c$").

$$K_f = P_f C^T \Theta^{-1} \quad (5)$$

$$P_f A^T + AP_f - P_f C^T \Theta^{-1} C P_f + \Gamma \Xi \Gamma = 0 \quad (6)$$

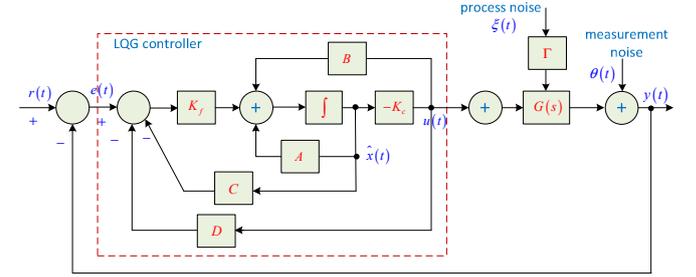

Fig. 1. LQG controller as combination of Kalman filtering based optimal state feedback controller.

As a next step of LQG design, estimated states $\hat{x}(t)$ are assumed to be a faithful representation of the original system states $x(t)$ in order to design an optimal state feedback law via LQR. The LQR design helps to keep the excursion of all the three state variables i.e. angle of attack, pitch deflection and canard's deflection, as low as possible since they are associated with some mechanical systems. It also takes the required control input $u(t)$ into account, producing an optimal control law minimizing the cost function given by (7).

$$\begin{aligned} J &= \lim_{t_f \to \infty} E \left\{ \int_0^{t_f} \begin{bmatrix} x^T & u^T \end{bmatrix} \begin{bmatrix} Q & N_c \\ N_c^T & R \end{bmatrix} \begin{bmatrix} x \\ u \end{bmatrix} dt \right\} \\ &= \lim_{t_f \to \infty} E \left\{ \int_0^{t_f} \left( x^T Q x + u^T R u \right) dt \right\} \quad [\because N_c = 0] \end{aligned} \quad (7)$$

Minimization of the cost function (7) turns out as finding symmetric positive semi-definite solution ($P_c = P_c^T$) of ARE (8) to obtain the optimal state feedback gain matrix $K_c$ in (9).

$$A^T P_c + P_c A - P_c B R^{-1} B^T P_c + Q = 0 \quad (8)$$

$$K_c = R^{-1} B^T P_c \Rightarrow u(t) = -K_c \hat{x}(t) \quad (9)$$

The two stage LQG control scheme of using the Kalman filter first for state estimation of a system with process/measurement noise and then using those estimated states for optimal LQR state feedback control is known as the separation principle. Details of the observer based LQG controller is shown in Fig. 1. The estimated states of the Kalman filter and also observer based LQG controller are given by (10) and (11) respectively. The observer based state-feedback controller i.e. LQG can be translated as an output feedback controller where the system description along with noise inputs are shown in Fig. 2 and the corresponding controller structure in Fig. 1.

$$\dot{\hat{x}} = A\hat{x} + Bu + K_f(y - C\hat{x} - Du) \quad (10)$$

$$G_c^{LQG} = K_c(sI - A + K_f C + BK_c - K_f DK_c)^{-1} K_f \quad (11)$$

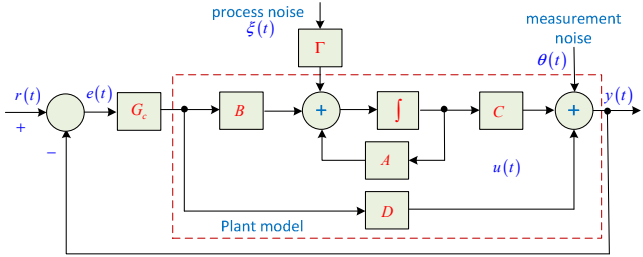

Fig. 2. Observer based LQG controller with external noise terms.

### C. LTR Technique to Increase Stability Margins

The well-known problem of the LQG control design is its reduced stability margin which may destabilize the system even with a small amount of disturbance. This is due to the fact that often the dynamics of the Kalman filter is not much faster compared to the plant dynamics. Considering the feed-through matrix $D = 0$ in the missile model (1), the open loop system with the LQG controller and plant can be represented as (12).

$$G_{ol}^{LQG} = K_c(sI - A + K_f C + BK_c)^{-1} K_f C(sI - A)^{-1} B \quad (12)$$

Since the controller structure involves Kalman filter gain $K_f$ and LQR state-feedback matrix $K_c$, wrong choice of the noise covariance matrices and weights may make the overall system have reduced stability margin or even lead to instability. The idea of LTR design is to use a fictitious-noise coefficient ($q$) along with the process noise covariance such that $\Xi' = q\Xi$ and then gradually increasing $q \to \infty$, so that the open loop system approaches to that of the LQR [22].

$$\underset{q \to \infty}{Lim} G_{ol}^{LQG} = \underset{q \to \infty}{Lim} K_c(sI - A + K_f C + BK_c)^{-1} K_f C(sI - A)^{-1} B$$
$$\simeq K_c(sI - A)^{-1} B = G_{ol}^{LQR} \quad (13)$$

The LTR controller design can be summarized in two steps:
- Choose suitable $\{Q, R\}$ to get a reasonably good LQR gain matrix $K_c$, such that the open-loop transfer function $-K_c(sI - A)^{-1} B$ meets desired performance measure like sensitivity, complementary sensitivity, gain and phase margins.
- Increase $q$ in $\Xi' = q\Xi$ and obtain $K_f$ using (5)-(6) until the solution approaches to LQR open loop performance following equation (12). Compute the LTR controller using the LQG controller structure (11) for the chosen $q$.

In practice, too large value of $q$ affects the robustness of the system and makes the initial control input very high. In LTR theory, there might be two options of recovering the input side (Kalman filter) or output side (LQR controller) of the system given by (14). However, the two step separation approach is popular among contemporary research community [19]-[20].

$$\underset{q \to \infty}{Lim} G_{ol}^{input} \simeq K_c(sI - A)^{-1} B$$
$$\underset{q \to \infty}{Lim} G_{ol}^{output} \simeq C(sI - A)^{-1} K_f \quad (14)$$

### D. LQI Control for Guaranteed Tracking Performance

In Huerta et al. [25] a modified LQG as a servo controller has been proposed along with an integrator for guaranteed tracking performance. The concept can be considered as somewhat similar to the LQI scheme for the deterministic case [26]-[27]. The schematic diagram of classical state-feedback control to LQG regulator and finally LQG servo control as a hybridization of LQG and LQI is shown in Fig. 3. The novel control scheme proposed in this paper first designs the LQG-LQI control scheme as in Fig. 3 and then by increasing $q \to \infty$, it brings the effect of LTR principles along with an optimization based tuning of LQR weights $\{Q, R\}$. In fact, within an optimisation framework and with a suitable initial guess of $\{Q, R, q\}$, the adopted hybrid LQG-LTR-LQI scheme can enjoy all the benefits of these three individual schemes.

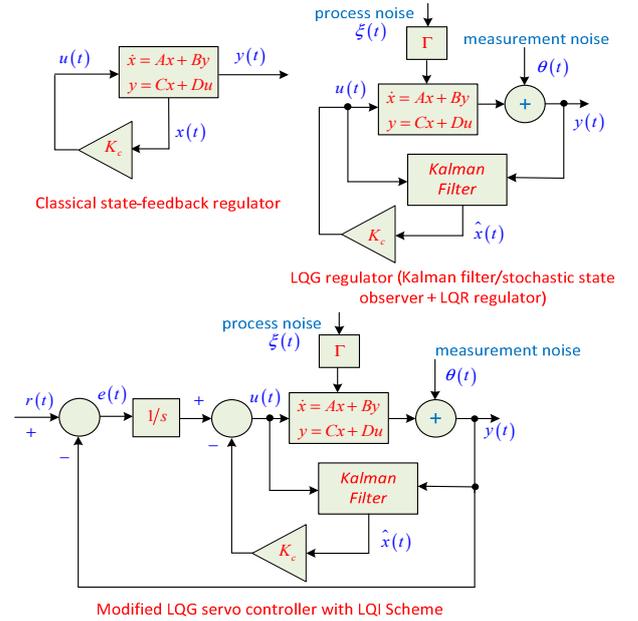

Fig. 3. Modified LQI scheme for LQG servo controller.

## III. SIMULATION STUDIES WITH LQG/LTR CONTROLLER

### A. Independence of Noise Covariances and LQR Weights

To study the parametric influence of the noise covariance matrices and LQR weights, for the nominal missile model (1), in the design phase they are considered as $\Xi = 10^{-3}$, $\Theta = 10^{-7}$,

$Q = 0.01 I_{3\times3}$, $R = 0.01$. The nominal system parameters are considered as $\Delta M_\alpha = 0, h = 0.5$ in all simulations. It is evident from Fig. 4 that a higher $h$ makes the system less oscillatory and the dynamics degrades to a greater extent with low $h$, compared to a wider variation in $\Delta M_\alpha$. The step response LQG control and effect of varying LQR weights ($Q, R$) are shown in Fig. 5. The impact of giving more penalties either on wider excursion of states or the controller effort is evident in Fig. 5 which follows an inverse relationship as conflicting objectives.

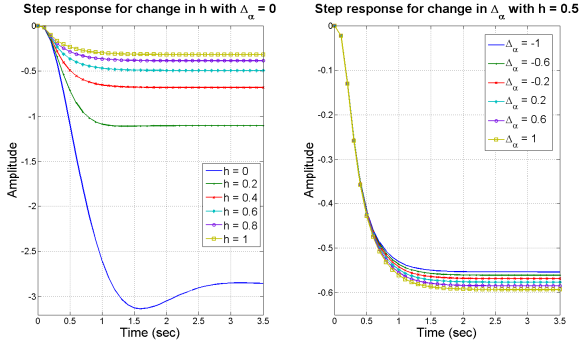

Fig. 4. Change in missile dynamics with variation in system parameters.

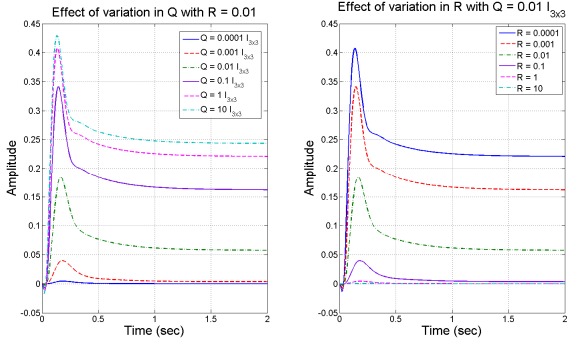

Fig. 5. Effect of varying LQG weighting matrices Q and R for noisefree case.

### B. Using Noise Covariances as LQR Weights in LQG/LTR

In the following simulations, the noise covariance matrices in Kalman filter design are considered as the LQR weights i.e. $Q = \Xi I_{3\times3}, R = \Theta$. Additionally, the effect of increasing the LQG/LTR fictitious noise coefficient $q$ for step response simulation study in noisy environment is also explored. In the LTR design, it is well known that increasing the value of $q$ pushes the Nyquist curve of the noise-free system to approach the LQR design [22]. Due to the presence of process noise and measurement noise terms, the frequency responses cannot be evaluated and therefore the focus here is on a realistic scenario of step input tracking of the LQG/LTR based missile attitude control system. Since with the LQR design the noise terms $\{\xi, \theta\}$ are not considered, therefore the time domain performance is expected to be worse than the LQG/LTR scheme. Fig. 6 shows that a high value of $q$ increases the steady-state value of the system output but makes the control signal highly oscillatory. The corresponding frequency responses in Fig. 7 show that the gain margin increases with high value of $q$. The only disadvantage of this scheme, as can be seen from Fig. 6 that the steady-state offset cannot be removed even with a wide variation in LQR weights.

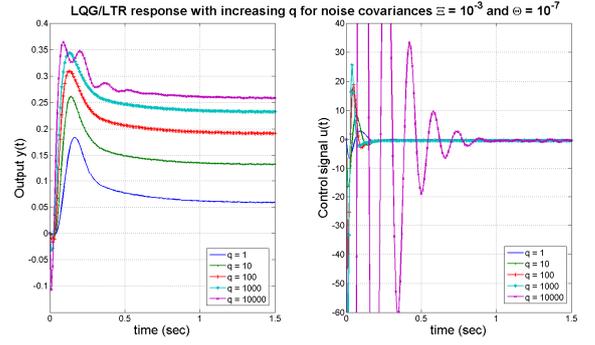

Fig. 6. System output and control signal for increasing $q$ in LTR.

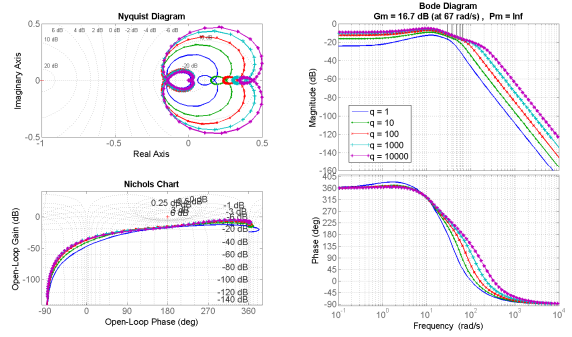

Fig. 7. Frequency response of the noise-free missile system with LTR.

### IV. HYBRID LQG-LTR-LQI CONTROL TO ENSURE SET-POINT TRACKING PERFORMANCE OF THE MISSILE

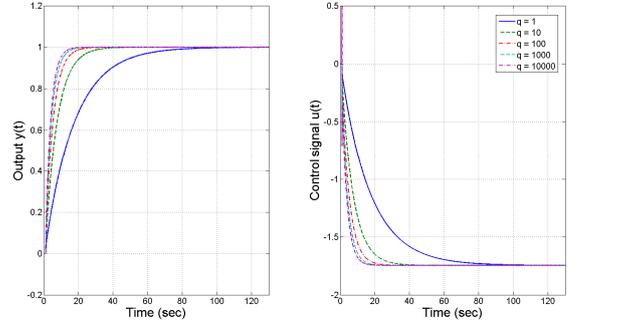

Fig. 8. Hybrid LQG-LTR-LQI response with a step input at $t = 1$ sec.

The LQG controller is optimal in a sense that given the noise covariance matrices, the Kalman filter produces optimal state estimates from noisy measurements and then the LQR enforces an optimal state feedback control law. But LQG suffers from reduced stability margin that can be improved using LTR by increasing the value of $q$ as shown in Fig. 7. The LQI scheme based control loop in Fig. 3 is explored next for guaranteed tracking. Simulations are shown in Fig. 8 for the hybrid LQG-LTR-LQI scheme with increasing $q$ and the nominal weighting matrices of LQR, as mentioned above. It is evident that for high value of $q$ in LQG-LTR-LQI controller, the initial controller effort at $t = 1$ sec suddenly becomes large which may cause actuator saturation. The scheme with integrator enforces the set-point tracking performance along with rejection of process noise and measurement noise externally introduced in the control loop. But arbitrary choice of $Q$ and $R$ has made the system somewhat sluggish. The problem may be circumvented by choosing optimal values of $Q$ and $R$ with an optimization algorithm, as discussed next.

### A. Finding Optimal Q and R for LQG-LTR-LQI Scheme

The well-known Nelder-Mead Simplex optimisation algorithm is employed to search for the LTR fictitious noise coefficient ($q$), diagonal elements of LQR weighting matrix ($Q1$, $Q2$, $Q3$) and LQR weighting factor ($R$) by minimising the Integral of Squared Error (ISE) criterion (15) for the closed loop system. The ISE can be easily evaluated from the step response of the sensitivity function $S(s)$ using available numerical integration techniques like trapezoidal rule.

$$\hat{J}_{ISE} = \int_0^\infty e^2(t)dt = \int_0^\infty \mathcal{L}^{-1}\left[\frac{1}{s}\frac{1}{1+G_{ol}(s)}\right]^2 dt = \int_0^\infty \mathcal{L}^{-1}\left[\frac{1}{s}S(s)\right]^2 dt \quad (15)$$

The control signal can be obtained from the inverse Laplace of the step input command to the control sensitivity function (16).

$$u(t) = \mathcal{L}^{-1}\left[(1/s)\left(G_c(s)/(1+G_{ol}(s))\right)\right] \quad (16)$$

In literature, the popular method is to set $Q = C^T C$ and varying $R$ to meet design specifications. The other approach includes simultaneous tuning of both $\{Q, R\}$ as in [23]-[24]. For the present study the following two cases are compared:

- Tuning all $Q, R, q$ simultaneously
- Assuming $Q = C^T C$ and then tuning only $R$ and $q$

TABLE I. OPTIMISATION RSULTS FOR ALL CONTROLLER PARAMETERS

| $\hat{J}_{min}$ | $q_0^{ini}$ | $q_{opt}$ | $Q_1$ | $Q_2$ | $Q_3$ | $R$ |
|---|---|---|---|---|---|---|
| 5.067 | 1 | 0.933 | 0.354 | 1.852 | 0.123 | 1.809 |
| 5.132 | 10 | 10.082 | 0.959 | 1.023 | 1.048 | 1.011 |
| *4.916* | 100 | 96.698 | 0.481 | 1.477 | 1.511 | 0.268 |
| 4.992 | 1000 | 1174.847 | 0.702 | 1.389 | 1.301 | 0.441 |

TABLE II. OPTIMISATION RESULTS OF THE PARAMETERS WITH $Q=C^T C$

| $\hat{J}_{min}$ | $q_0^{ini}$ | $q_{opt}$ | $R$ |
|---|---|---|---|
| 5.777 | 1 | 1.250 | 1.300 |
| 5.850 | 10 | 30.869 | 1.123 |
| *5.743* | 100 | 100.036 | 1.474 |
| 5.811 | 1000 | 1024.804 | 1.248 |

The Nelder-mead simplex optimisation algorithm runs with an initial guess and then it converges to the minima using an iterative method. The initial guess for the $\{Q, R\}$ are set to unity for all cases, except for the LTR where the initial value of fictitious noise coefficient in the Kalman filter design is chosen as $q_0^{ini} \in \{1, 10, 100, 1000\}$. For the two cases, as mentioned above, the optimisation results are shown in Table I and II respectively where it shows that in both cases the $q_0^{ini} = 100$ gives the best tracking performance, judged by the minimum ISE or $\hat{J}_{min}$. The effect of increase in $q$ shows improvement in gain margin in Fig. 9-10, for the two cases of optimization. However, the practically implementable values should be judged from the tracking performance and required controller effort which has turned out as the corresponding optimal weights in the third row of both Table I-II. Also the advantages of tuning all the parameters rather than tuning only $R, q_0$ is evident from Fig. 9-10 and Table I-II, as the former produces consistently better tracking and stability margins.

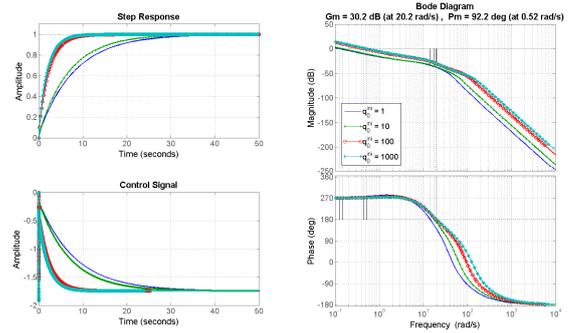

Fig. 9. Response with tuned weights $Q$, $R$, $q_0$ corresponding to Table I.

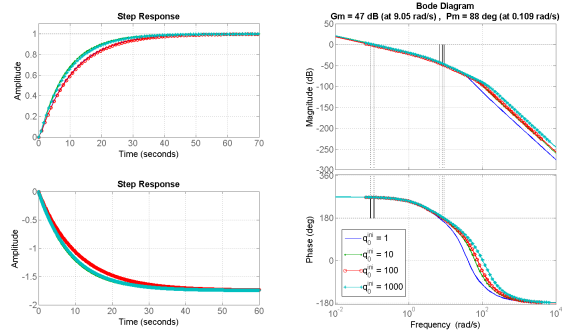

Fig. 10. Response with tuned values of $R$, $q_0$ corresponding to Table II.

### B. Effect of Wrong Estimate of Noise Covariances

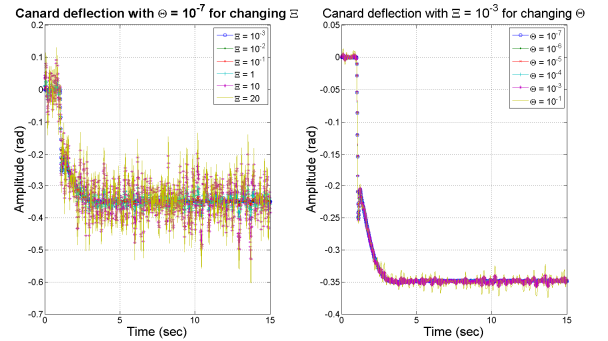

Fig. 11. Canard deflection for gradual increase in $\Xi$ and $\Theta$.

The best control scheme i.e. the proposed LQG-LTR-LQI scheme with $Q_{opt} = diag([1.477, 1.511, 0.268])$, $R_{opt} = 0.268$, $q_{opt} = 96.698$ has now to be tested with a wrong estimate of the noise covariance matrix. The resulted transfer function of the hybrid optimal controller for the present canard missile attitude control problem has been computed as (17).

$$G_c^{opt}(s) = \frac{-548.94(s^2 + 17.8s + 159.1)}{s(s + 39.95)(s^2 + 31.48s + 1090)} \quad (17)$$

In a practical scenario, the noise covariance must be continuously updated affecting the Kalman gain in (5). The LQG controller should also be updated for the best noise rejection performance. Here in Fig. 11, the effects of increasing the noise covariance matrices on the system output, control input and canard deflection are studied, keeping the controller tuned at a lower noise specification $\Xi = 10^{-3}, \Theta = 10^{-7}$. It is evident that the scheme is capable of keeping the canard deflection at lower level for a range of noise characteristics.

## V. CONCLUSION

An efficient attitude control system for the canard missile has been proposed in this paper as a fusion LQG, LTR and LQI techniques with an optimization based weight selection. The controller guarantees set-point tracking, rejects process noise and measurement noise and also provides better stability margins than that can be achieved from each of the techniques separately. The optimization procedure reduces the manual heuristics needed while tuning of the proposed LQG-LTR-LQI controller. Future work may be directed towards extending the concept for correlated noise terms and nonlinear missile model.